\documentclass[11pt,twoside]{article}

\usepackage{asp2004}
\usepackage{epsf}
\usepackage{psfig}
\usepackage{lscape}
\usepackage{amssymb}

\newcommand{\rh}{\mbox{$r_{\rm h}$}}

\newcommand{\Mc}{\mbox{$M_{\rm c}$}}
\newcommand{\Ec}{\mbox{$E_{\rm c}$}}
\newcommand{\de}{\mbox{${\rm \Delta} E/|E_{\rm c}|$}}
\newcommand{\dm}{\mbox{${\rm \Delta} M/M_{\rm c}$}}
\newcommand{\tdis}{\mbox{$t_{\rm dis}$}}
\newcommand{\rhon}{\mbox{$\rho_{\rm GMC}$}}
\newcommand{\rhoc}{\mbox{$\rho_{\rm c}$}}
\newcommand{\rrms}{\mbox{$\bar{r^2}$}}
\newcommand{\gm}{\mbox{$g_{\rm m}$}}
\newcommand{\dr}{\mbox{${\mbox d}$}}
\newcommand{\at}{\mbox{$a_{\rm t}$}}

\newcommand{\dv}{\mbox{$\Delta v_{\rm x}$}}
\newcommand{\Mdot}{\mbox{$\dot{M_{\rm c}}$}}

\begin{document}
\title{The role of tidal forces in star cluster disruption}    
\author{M. Gieles}   
\affil{Astronomical Institute, Utrecht University, Princetonplein 5, 3584-CC Utrecht, The Netherlands}    

\begin{abstract} 
Star clusters are subject to density irregularities in their host
galaxy, such as giant molecular clouds (GMCs), the galactic
disc and spiral arms, which are largely ignored in present day
($N$-body) simulations of cluster evolution. Time dependent external
potentials give rise to tidal forces that accelerate stars leading to
an expansion and more rapid dissolution of the cluster. I explain the
basic principles of this tidal heating in the {\it impulse
approximation} and show how related disruption time-scales depend on
properties of the cluster.
\end{abstract}

\section{Tidal heating, how does it work?}
Assume a cluster and a moving perturbing object (GMC, spiral arm,
disc, etc.) with relative separation and velocity $X$ and $V$,
respectively. The acceleration due to the perturber is referred to as
$g(X)$. The {\it tidal} acceleration ($\at$) on a star at a distance
$x$ from the cluster centre can be obtained by linearising $g(X)$:
\begin{equation}
\at=\frac{\dr v_{\rm x}}{\dr t} = g(X+x)-g(X)\simeq x\,\frac{\dr g}{\dr X}.
\label{eq:1}
\end{equation}
The total velocity increase ($\dv$) of the star after the perturbation
can be obtained by substituting $\dr X = V \dr t$ and integrating
Eq.~\ref{eq:1}. If the time it takes the perturber to pass the cluster is
much shorter than the crossing time of stars in the cluster,
$x$ can be assumed constant. This is referred to as the {\it impulsive
approximation}. The resulting \dv\ then simply scales with $x$, $V$
and the maximum acceleration ($\gm$) as: $\dv \propto x\,\gm/V$.  The
total kinetic energy gain per unit of cluster mass ($\Delta E$) can
then be written as
\begin{equation}
\Delta E = \frac{1}{2}|\dv^2| \propto \frac{\gm^2\,\rrms}{V^2},
\label{eq:2}
\end{equation}
where we used that the mean-square position of stars (\rrms) is
$\rrms=3\bar{x^2}$. {\it Eq.~\ref{eq:2} shows that $\Delta E$ is
 large for slow perturbations on large clusters}.

For passing point masses $\gm=GM/p^2$, with $G$ the gravitational
constant, $M$ the mass of the perturber and $p$ the distance of
closest approach (see \citealt{2006MNRAS.tmp..808G} for the exact
definition of Eq.~\ref{eq:2}, taking into account an extended density
profile for a GMC). For a one-dimensional density perturbation, such
as the perpendicular crossing of a galactic disc and, to lesser
extend, the crossing of a spiral arm, $\gm \propto GH\rho_0$, with $H$
the scale height/length and $\rho_0$ the central density. For more
details see \citet{1987degc.book.....S} and references therein.

\section{From energy gain to mass loss}
A single tidal perturbation can accelerate stars to velocities much higher than
the escape velocity. This {\it overkill} energy can not be used to
unbind more stars and is lost with the escaping stars. For this
reason, the fractional mass loss (\dm, with $\Mc$ the cluster mass) is
lower than the fractional energy gain (\de, with $\Delta E$ from
Eq.~\ref{eq:2} and $\Ec$ the cluster energy per unit mass), by a
factor $f$, with $3\lesssim f\lesssim5$
\citep{gieles06a,2006MNRAS.tmp..808G}. With this and the definition of
\Ec\ ($\Ec=\eta G\Mc/\rh$, with \rh\ the cluster half-mass radius and
$\eta\simeq0.2$, depending on the cluster density profile), we can
write $\Delta M$ as

\begin{equation}
\Delta M = \frac{\Mc}{f}\frac{\Delta E}{|\Ec|}\propto\Mc\rrms\frac{\rh}{\Mc}\propto\rh^{3},
\end{equation}
where we used Eq.~\ref{eq:2} and $\rrms=C\rh^2$, with $1.5\lesssim
C\lesssim3.5$, depending on the cluster density profile. The
statistical mean mass loss per unit time ($\Mdot$) is defined as
$\Mdot\equiv \Delta M/\Delta t$, where $\Delta t$ for GMC encounters
is the time between successive encounters and depends on the GMC
density ($\rhon$) as $\Delta t \propto \rhon^{-1}$. {\it Note that $\Delta
M$ and $\Mdot$ due to external perturbations are independent of
cluster mass.}

\section{From mass loss to disruption time}
Defining the cluster disruption time ($\tdis$) as $\tdis=\Mc/\Mdot$
\citep{2005A&A...441..117L}, we can write (for GMC encounters)

\begin{equation}
\tdis=\Mc/\Mdot \propto \frac{\Mc}{\rh^3}\rhon^{-1}\propto\rhoc\,\rhon^{-1}.
\label{eq:4}
\end{equation}
From observations of clusters in different galaxies,
\citet{2003MNRAS.338..717B} found that $\tdis\propto\Mc^{\gamma}$,
with $\gamma\simeq0.6$. The same value for $\gamma$ was found in
$N$-body simulations of clusters dissolving in a galactic tidal field
\citep{2003MNRAS.340..227B}, however, their \tdis\ values are much
longer than the observed ones. 
\citet{2004A&A...416..537L} recently
found that $\rh\propto\Mc^{0.1}$, which implies
$\rhoc\propto\Mc^{0.7}$ and hence $\tdis\propto\Mc^{0.7}$
(Eq.~\ref{eq:4}).  The fact that \tdis\ due to the Galactic tidal
field {\it and} due to the tidal perturbations depend on \Mc\ in
almost the same way can explain why the values of $\gamma$ derived
from observations of clusters in different galaxies are so similar
\citep{2003MNRAS.338..717B}, while \tdis\ varies by almost two
orders of magnitude \citep*{2005A&A...429..173L}.

\end{document}